# Revealing the nature of the ultrafast magnetic phase transition in Ni by correlating extreme ultraviolet magneto-optic and photoemission spectroscopies

Wenjing You[1†], Phoebe Tengdin[1†], Cong Chen[1], Xun Shi[1]*, Dmitriy Zusin[1], Yingchao Zhang[1], Christian Gentry[1], Adam Blonsky[1], Mark Keller[2], Peter M. Oppeneer[3], Henry Kapteyn[1], Zhensheng Tao[1*‡], Margaret Murnane[1]

[1]*Department of Physics and JILA, University of Colorado and NIST, Boulder, Colorado 80309, United States*
[2]*National Institute of Standards and Technology (NIST), 325 Broadway, Boulder, Colorado 80305, United States*
[3]*Department of Physics and Astronomy, Uppsala University, Box 516, 75120 Uppsala, Sweden*

[†]These authors contributed equally to this work.

*Corresponding authors: Dr. Xun Shi, xun.shi@colorado.edu,

Dr. Zhensheng Tao, zhensheng.tao@jila.colorado.edu.

## Abstract

By correlating time- and angle-resolved photoemission and time-resolved transverse- magneto-optical Kerr effect measurements, both at extreme ultraviolet wavelengths, we uncover the universal nature of the ultrafast photoinduced magnetic phase transition in Ni. This allows us to explain the ultrafast magnetic response of Ni at all laser fluences - from a small reduction of the magnetization at low laser fluences, to complete quenching at high laser fluences. Both probe methods exhibit the same demagnetization and recovery timescales. We further show that the ultrafast demagnetization in Ni is indeed a magnetic phase transition that is launched within 20 fs, followed by demagnetization of the material within ≈200 fs, and subsequent recovery of the magnetization on timescales ranging from 500 fs to >70 ps. We also provide evidence of two competing channels with two distinct timescales in the recovery process, that suggest the presence of coexisting phases in the material.

‡ Current address: State Key Laboratory of Surface Physics, Department of Physics, Fudan University, Shanghai 200438, People's Republic of China

Magnetization in magnetic materials can be strongly suppressed by ultrafast laser irradiation on femtosecond timescales [1]. Numerous experiments have been performed on transition-metal ferromagnets (Co, Ni, and Fe) to show that the magnetization is quenched within ~100 to 500 fs, before subsequently recovering within tens of picoseconds [2–11]. More recently, all-optical control of the magnetic state of a material has attracted great attention, having been recently realized in ferrimagnetic alloys [12], ferromagnetic multilayers [13] and other compounds [14]. Understanding the microscopic mechanisms underlying fast spin manipulation is of fundamental interest and also has implications for future data-storage and spintronic devices. As a result, ultrafast magnetic phase transitions have been studied using many experimental techniques, including magneto-optical spectroscopy [3,5–7,9,11], photoelectron spectroscopy [2,4,8], and X-ray magnetic circular dichroism [10,15].

Despite these extensive experimental efforts, the underlying physical mechanisms that drive ultrafast magnetization dynamics are still under debate. A number of microscopic models based on mechanisms such as Elliott-Yafet spin-flip scattering [5,16], dynamic exchange splitting reduction [17–19], as well as ultrafast spin-polarized or unpolarized currents [20,21], have been proposed. In addition, coherent optical excitation [22], spin-orbit coupling [23,24] and collective magnon excitation [3,25,26] are also believed to play an important role in this process. In the past, the difficulty in determining the correct underlying mechanism was due to several issues: First, standard magneto-optic spectroscopies are simply not sensitive to highly non-equilibrium excited magnetic states, without simultaneously monitoring the coupled electron, spin, and lattice degrees of freedom. Second, these spectroscopies average over different depths of the material, which masked the physics of the ultrafast phase transition.

In recent work, using time- and angle-resolved photoemission (Tr-ARPES), we unambiguously revealed the existence of critical phenomena during ultrafast demagnetization in Ni. Specifically, we uncovered the existence of a critical laser fluence, above which the electron temperature is driven above the Curie temperature, and the material subsequently undergoes a magnetic phase transition [27]. Given this new understanding of the importance of critical phenomena in ultrafast magnetic phase transitions, it is now essential to revisit results obtained using magneto-optical techniques, to understand how to interpret them correctly.

In this work, we investigate the ultrafast magnetic phase transition in Ni using time-resolved transverse- magneto-optical Kerr effect (Tr-TMOKE) spectroscopy based on high harmonic generation. Using the critical behavior and the timescales of demagnetization and recovery processes observed from Tr-ARPES, and by taking the depth-dependent signal contributions in Tr-TMOKE into account, we show that critical phenomena are also key for the correct interpretation and a full understanding of optical/X-ray magnetic spectroscopies. With this knowledge, we can now fully explain the Tr-TMOKE response of Ni over the full range of laser fluences, using only three universal timescales to describe the demagnetization and recovery dynamics in distinct physical regions. While the spin system is excited within ~ 20 fs after the peak of the driving laser pulse, the spectroscopic signatures of demagnetization take ~ 176 fs to develop. Moreover, both of these timescales are fluence-independent. In contrast, the speed of re-magnetization dynamics depends on whether the applied laser fluence is below or above the critical fluence (see Fig. 1). Our data show that the demagnetization amplitudes scale linearly with pump fluence. Finally, we observe a competition between the fast and slow recovery channels with distinct timescales, suggesting a potential coexistence of ferromagnetic and paramagnetic phases during the phase transition.

We note that the ability to manipulate the magnetic state on femtosecond timescales is important both scientifically and technologically. Although ferromagnetic metals are some of the simplest materials that exhibit strong interactions between the electron, spin and lattice degrees of freedom, there is yet no comprehensive theory that describes their non-equilibrium behavior. Past work concluded that many different timescales were associated with laser-induced magnetic dynamics, and that these depended on the pump fluence [16,28] and sample geometry [29,30]. This made it challenging to develop complete theories and compare with experiments. In contrast, by showing the essential contribution of critical behavior associated with a magnetic phase transition, we reveal that only a few characteristic timescales are needed to fully explain ultrafast demagnetization in Ni.

A schematic of the experimental setup is shown in Fig. 1a. The sample used in our experiments was a 400 nm Ni(111) single-crystalline film. We intentionally chose a thick film sample to minimize nonlocal effects due to interfaces or poor substrate thermal conduction [29,30] and also verified that the observed dynamics were not dependent on the orientation of the sample (see Supplemental Material (SM)). In both the Tr-TMOKE and Tr-ARPES experiments, the sample was excited by ~ 45 fs pulses from a Ti:Sapphire laser amplifier system at a wavelength of 800 nm. In the Tr-TMOKE measurements, the subsequent change of the sample magnetization was probed by extreme ultraviolet (EUV) pulses produced by high harmonic generation (HHG). The sample magnetization can be quantitatively determined by recording the asymmetry of the reflected HHG spectrum at the 3*p* edge of Ni [6,7,11]. In the Tr-ARPES measurements, the magnetization dynamics was probed by monitoring the magnitude of the exchange splitting at different time delays [2,8,27].

In order to determine if Tr-TMOKE and Tr-ARPES give spectroscopic signatures that are consistent with the same microscopic physics and interactions, we measured the de- and re-magnetization dynamics in Ni excited by a wide range of fluences, with the highest fluence sufficient to fully suppress the Tr-TMOKE asymmetry (i.e., demagnetize the sample). The pump penetration depth in Ni is $\delta_L$ ~ 13 nm [31], which is comparable to the probing depth of the EUV light used in the Tr-TMOKE experiments (~ 10 nm). In contrast, the probing depth of photoelectrons is close to a monolayer for the photon energy (~ 16 eV) used in the Tr-ARPES experiments [32], which suggests that the Tr-ARPES signal can probe the elementary magnetization dynamics in an individual surface layer of the sample. In Fig. 1b, we conceptually summarize the electron, spin and magnetization dynamics after laser excitation, with the critical behavior taken into consideration [27].

In Fig. 2, we plot the change of the exchange splitting ($\Delta E_{ex}$) at the transverse momentum $k_{//}$ ≈ 1.05 Å$^{-1}$ along the $\bar{\Gamma}-\bar{\mathrm{K}}$ direction of Ni (inset of Fig. 2) observed in the Tr-ARPES measurements [27]. Due to <1nm probing depth, Tr-ARPES probes the elementary magnetization dynamics in a monolayer of the material, which can be well described by an exponential decay and bi-exponential recovery function as shown in Fig. 2:

$$m(t_d, z) = \begin{cases} 1 & (t_d < 0) \\ 1 + a_1(z) e^{-\frac{t_d}{\tau_{\mathrm{demag}}}} - a_2(z) e^{-\frac{t_d}{\tau_{\mathrm{recover1}}}} - a_3(z) e^{-\frac{t_d}{\tau_{\mathrm{recover2}}}} & (t_d \geq 0) \end{cases} \quad (1)$$

Here we obtain three time constants that correspond to the following physical processes: the collapse of the exchange splitting $\tau_{\mathrm{demag}}$ =176 ± 27 fs; a fast recovery time $\tau_{\mathrm{recover1}}$ = 537 ± 173 fs; and a slow recovery time $\tau_{\mathrm{recover2}}$ =76 ± 15 ps [27]. See SM for data supporting the extraction of the time constants. In Eq. (1), $a_1$, $a_2$ and $a_3$ are the amplitudes of these processes, with $a_1$=$a_2$+$a_3$.

Note that only two of the amplitudes are independent since the magnetization will recover fully at long times. Their values depend on the strength of the laser fluence, and, hence, are depth dependent due to the profile of the optical pump below the sample surface (Fig. 1a). From the ARPES results, we map the dynamics in monolayers of the material - we can now test whether this understanding can fully explain the Tr-TMOKE results.

The magnetization dynamics in the same sample excited by fs laser irradiation were also measured using Tr-TMOKE. In the inset of Fig. 3, we present the bulk-averaged amplitudes of de- and re-magnetization ($\langle A_1 \rangle$, $\langle A_2 \rangle$ and $\langle A_3 \rangle$) as a function of pump fluences, by fitting the Tr-TMOKE results presented in Fig. 3 with the same exponential decay and bi-exponential recovery function. Here the amplitudes represent the change of the sample magnetization normalized to the magnetization of the ground state. From these results, we find that the slow-recovery process ($\langle A_3 \rangle$) only turns on when the absorbed laser fluence is above the critical fluence ($F_c \approx 0.59$ mJ/cm$^2$), which highlights the importance of the critical behavior to the interpretation of the Tr-TMOKE results (Note that in [27] we quoted the incident fluence of 2.8 mJ/cm$^2$, which is consistent with an absorbed fluence of 0.59 mJ/cm$^2$ within error bars). Moreover, a linear response of the slow-recovery amplitude $\langle A_3 \rangle$ can be clearly observed, as highlighted in the inset of Fig. 3.

Under the assumption of linear absorption, the *in-situ* laser fluence $F$ decays exponentially with the depth $z$, i.e., $F(z) = F_0\, e^{-z/\delta_L}$ , where $F_0$ is the fluence at the surface. To take into account the true absorption at different depths, the heat source $q$ can be calculated by $q(z) = F(z)/\delta_L$ (see SM). When $F_0 > F_c$, the Tr-TMOKE signal arises from different regions, each exhibiting different recovery dynamics depending on whether the laser excitation is above or below the critical fluence (Fig. 1a). In *Region (i)* where the *in-situ* fluence is always above the critical fluence, the sample re-magnetizes through both slow and fast recovery channels. In contrast, in *Region (ii),* the *in-situ* fluence is lower than $F_c$, and re-magnetization occurs only through the fast channel. Here, we further assume that the change of magnetization is a linear function of the *in-situ* fluence, which is strongly supported by our experimental results (inset of Fig. 3) and previous work [33]. Given this linear relation, we have

$$a_1(z) = \min\left[b_1\, F(z), 1\right] \tag{2}$$

and

$$a_3(z) = \begin{cases} 0 & [F(z) < F_c] \\ \min\{b_3[F(z) - F_c], 1\} & [F(z) \geq F_c] \end{cases}, \tag{3}$$

where $b_1$ and $b_3$ are the proportionality constants. The Tr-TMOKE signals can be modeled as the bulk-averaged magnetization $\langle M \rangle$, given by the integral of the unit magnetization $m(t_d, z)$ over the probed depth $z$:

$$\langle M \rangle(t_d) = \frac{\int_0^\infty m(t_d, z) W(z)\, \mathrm{d}z}{\int_0^\infty W(z)\, \mathrm{d}z}. \tag{4}$$

Here $W(z)$ is the depth sensitivity function of TMOKE [34] , which is explicitly calculated for Ni (see SM for details).

Using the model described above, we now fit the Tr-TMOKE results for the different fluences shown in Fig. 3 to Eqs. (1-4), taking only $b_1$, $b_3$ and $F_c$ as the fitting parameters. We use the characteristic times obtained from the Tr-ARPES measurements as the time constants in Eq. (1) (see SM). As shown in Fig. 3, there is excellent agreement between the model (solid lines) and experimental data (symbols) over the full range of pump fluences, even though the limited number of fitting parameters places a strong constraint on our fitting. We note that the extracted value of $F_c$ is in good agreement with values obtained from the Tr-ARPES experiments [27], which further validates our model. From these results, we find the apparent presence of a fluence-dependent re-magnetization time is a direct result of the bulk-averaged signal in Tr-TMOKE: the surface layers of the material undergo a phase transition and exhibit slow recovery dynamics, while layers deeper within the material do not undergo a magnetic phase transition and as a result, exhibit only fast recovery dynamics. We note that similar fluence-dependent re-magnetization times have been often observed in previous Tr-TMOKE experiments on ferromagnets - these were interpreted as a frustration-induced slow-down of the spin dynamics [28], and were regarded as important evidence supporting the Elliott-Yafet spin-phonon interaction as the relevant microscopic mechanism [9,16]. In contrast, our model provides an alternative interpretation validated over the full demagnetization parameter space: there indeed exists a transient magnetic phase transition in Ni when the excitation laser fluence is higher than a critical value, which can completely explain the observed Tr-TMOKE data. The optimum values of fitting parameters are listed in Table 1.

From our model which correlates the Tr-TMOKE and Tr-ARPES results, we can extract the time- and depth-dependent magnetization dynamics in Ni. In Fig. 4a, we plot the amplitudes of the exponential functions in Eq. (1) for a monolayer Ni as a function of the heat source. A complete temporal and spatial profile of the laser-induced ultrafast demagnetization in Ni is plotted in Fig. 4b. Physically, the characteristic fast and slow recovery timescales ($\tau_{\mathrm{recover1}}$ and $\tau_{\mathrm{recover2}}$) indicate the existence of two distinct physical mechanisms. The fast re-magnetization timescale ($\tau_{\mathrm{recover1}}$) can be explained by damping of magnons under the strong exchange field in Ni [28], which yields a damping time of ~580 fs (see SM) , in quantitative agreement with the observed fast recovery timescale ($\tau_{\mathrm{recover1}}$, within experimental error) [27]. On the other hand, from molecular field theory, the exchange field is dissolved when the sample crosses the critical point and enters the paramagnetic state. In this case, we can expect the damping time to approach infinity and cooling of the spin system can only be achieved via other mechanisms, e.g., coupling to the lattice and thermal transport. The latter is consistent with the appearance of the slow re-magnetization process ($\tau_{recover2}$), when the fluence is above the critical fluence. As a result, the distinct timescales in our ultrafast measurement provide a way to probe the exchange field present on microscopic scales. Our results, hence, suggest the competition and coexistence of paramagnetic (slow recovery) and partially suppressed ferromagnetic (fast recovery) phases during the ultrafast demagnetization process, as well as the variation of their relative contributions as a function of pump fluence (Fig. 4c). Indeed, it has been shown by simulations based on atomic level classical spin Hamiltonian that the recovery from a highly disordered magnetic state involves the growth of many small magnetically ordered and disordered regions, with a size comparable to the magnetic correlation length [28].

Very interestingly, the fluence for which the fast-re-magnetization contribution completely disappears ($F_c$' in Fig. 4a), coincides with the fluence that drives the lattice temperature above the Curie temperature (see SM). This is consistent with the thermodynamic limit. We note, however, that we cannot simply conclude that the variation of sample magnetization is only determined by the electron/lattice temperature. One obvious evidence is that the magnetization at long delay times ($a_3$) increases linearly as a function of the laser fluence (and, hence, of the temperature), as shown in Fig. 4a - this cannot be explained by the typical nonlinear relationship between the sample magnetization and temperature under thermal equilibrium conditions (see SM). This result suggests that the spin system is far from thermal equilibrium on timescales of picoseconds, a finding which is consistent with previous theory [28]. By separating the different degrees of freedom in the time domain, our results suggest that the single critical point under thermal equilibrium is expanded into a critical region for the non-equilibrium magnetic phase transition in Ni (Fig. 4a), spanning critical fluences that first drive the electron temperature above the Curie temperature ($F_c$) and then the lattice to the Curie temperature ($F_c$').

Finally, another interesting conclusion we can make from our work is how to achieve very fast all-optical manipulation of spins, which has been an important goal ever since the first observation of ultrafast demagnetization [1]. In numerous previous experiments [1–11], it has been shown that the fundamental speed of this process is limited by the slow recovery dynamics, which typically occur on picosecond-to-nanosecond timescales. In this work, we provide a clear physical interpretation for this process: first, a magnetic phase transition is induced when the laser fluence is higher than the critical fluence ($F_c$). Then, for fluences between $F_c$ and $F_c$', the excited spin dynamics must then recover through a slow channel, likely restoring the magnetization through a combination of spin-lattice interactions and thermal transport. From our data, one way to achieve faster all-optical spin control on sub-ps timescales is to apply a laser fluence lower than $F_c$ - although in this case, the maximum demagnetization is < 50% in Ni, as shown in Fig. 4a. Another alternative would be to use a nanostructured material, with adjustable magnetic interactions and more optimal thermal transport.

In conclusion, we show that by correlating Tr-ARPES and Tr-TMOKE measurements on Ni, we obtain new insights into the laser-induced magnetic phase transition. All results consistently reveal a critical behavior associated with a true magnetic phase transition, and universal timescales for spin excitation, demagnetization, and recovery. Moreover, the linear response and two competing channels observed in the recovery process suggest the possible presence of co-existing phases in the material.

**Acknowledgements**

The experiments were performed at JILA. We gratefully acknowledge support from the Department of Energy Office of Basic Energy Sciences X-Ray Scattering Program Award DE-SC0002002 for supporting the magnetic TMOKE spectroscopy measurements performed for this work. We also thank the National Science Foundation through the JILA Physics Frontiers Center PHY-1125844, and the Gordon and Betty Moore Foundation EPiQS Award GBMF4538, for support of the ARPES measurements performed here. P.M.O. acknowledges support from the Swedish Research Council (VR), the Wallenberg Foundation (grant No. 2015.0060) and EU H2020 Grant No. 737709 "FEMTOTERABYTE". H.K. and M.M. have a financial interest in a laser company, KMLabs, that produces the lasers and HHG sources used in this work. H.K. is partially employed by KMLabs.

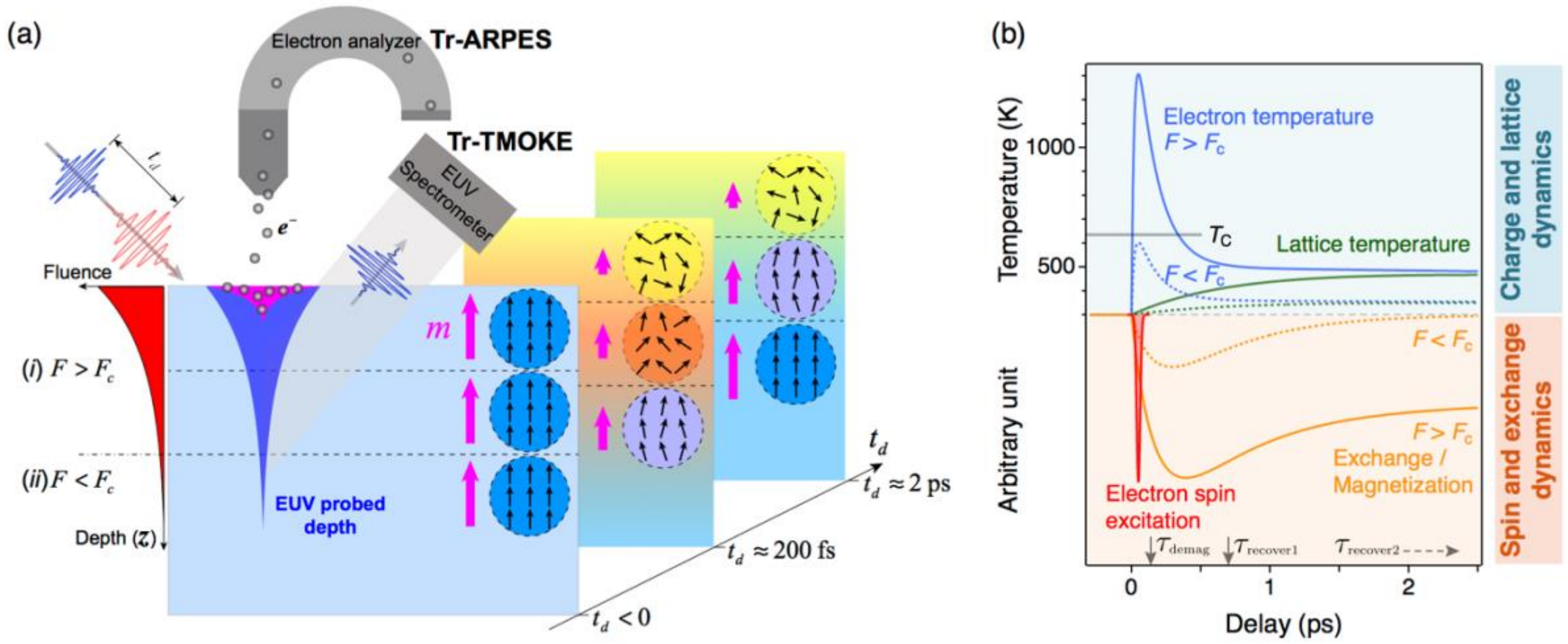

**Figure 1.** (a) Schematic of EUV ARPES and TMOKE measurements on Ni(111). The fluence profile of the laser excitation below the sample surface separates the magnetization response into two different regions (i) and (ii), depending on whether the *in-situ* fluence is above the critical fluence $F_c$. Using Tr-ARPES, the probed depth is on order of a monolayer, while Tr-TMOKE probes the entire laser-heated depth of ≈ 10nm. (b) Schematic of the excitation present in the laser-induced phase transition in Ni when critical phenomena are taken into consideration [27]. When the laser fluence exceeds the critical fluence $F_c$, the electron temperature exceeds $T_c$ and the sample rapidly undergoes a magnetic phase transition, as evidenced by multiple critical phenomena.

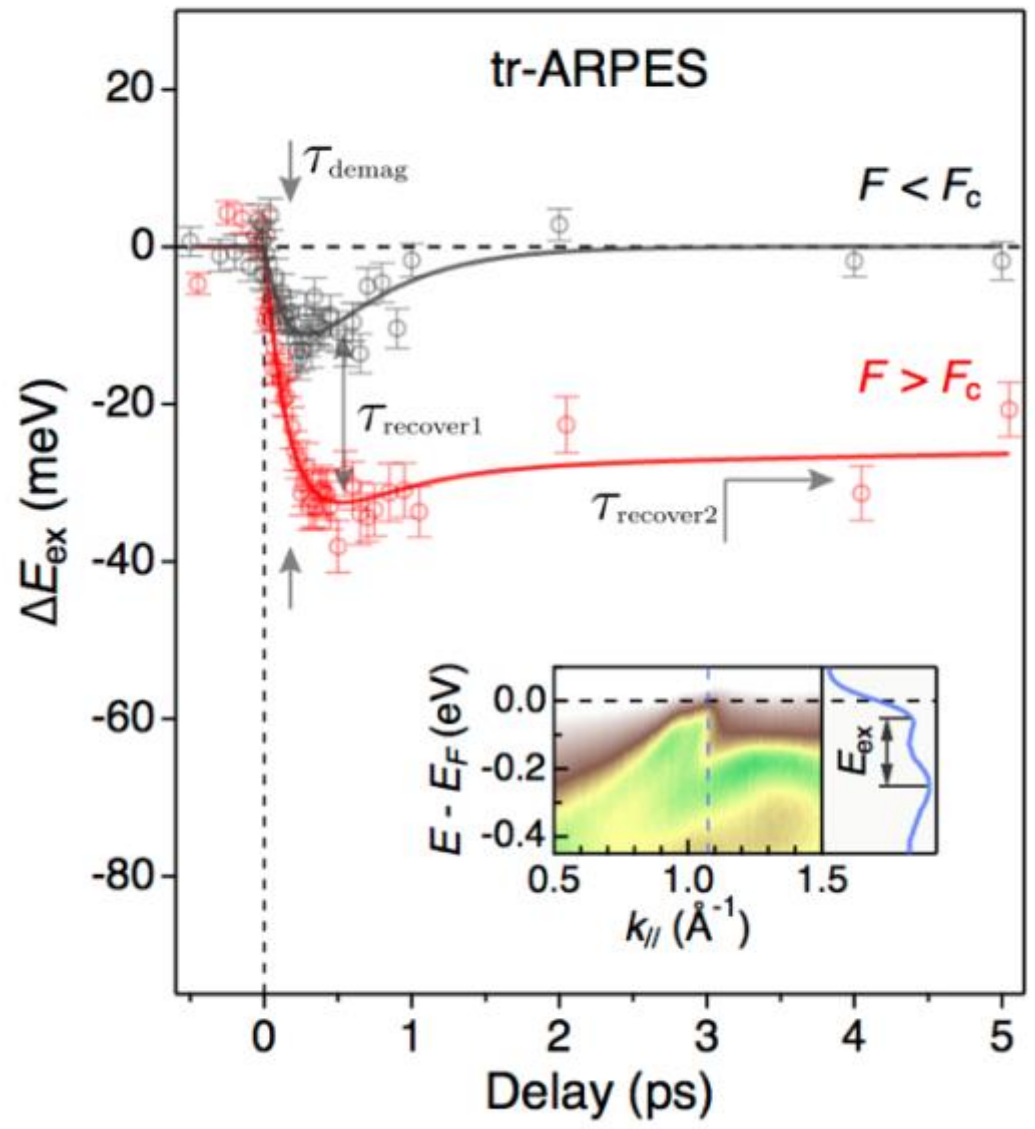

**Figure 2.** Change in the exchange splitting ($\Delta E_{ex}$) in Ni measured using Tr-ARPES, for the absorbed laser fluence below (0.21 mJ/cm$^2$, grey) and above (1.7 mJ/cm$^2$, red) the critical fluence $F_c$. The solid lines are the fits to Eq. (1). Inset: Static ARPES spectrum plot along the $\bar{\Gamma}-\bar{K}$ direction recorded using He I$_a$ photons.

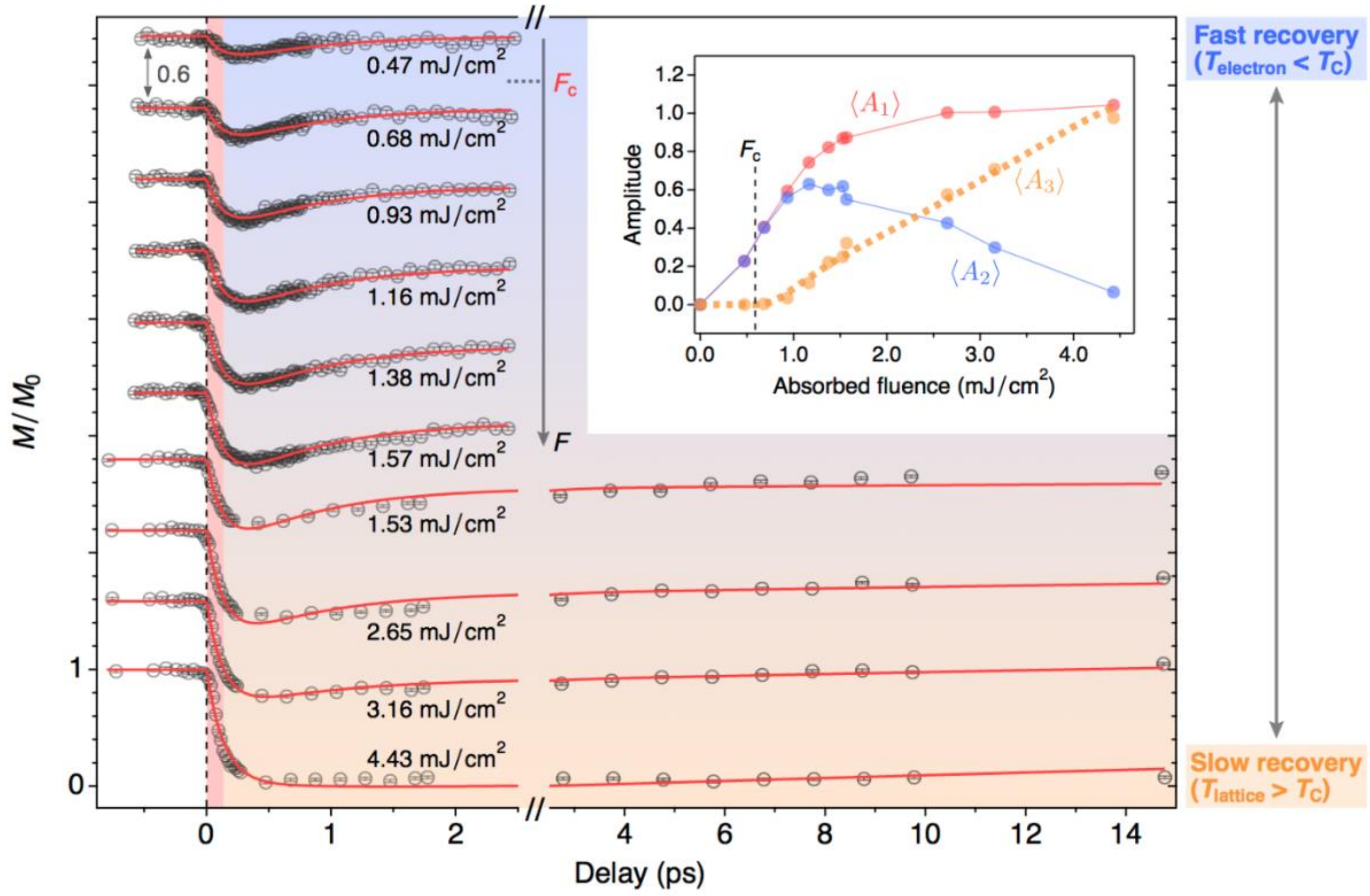

**Figure 3.** Magnetization dynamics in Ni measured using Tr-TMOKE over a full range of laser fluences. The highest fluence is sufficient to fully suppress the sample magnetization. The data are offset for clarity. Red curves: fits to our microscopic model which considers the critical behavior, as well as the depth-average effects in the Tr-TMOKE measurements. Inset: Fluence-dependent amplitudes of the demagnetization and recovery processes directly extracted from the Tr-TMOKE results. In the Tr-TMOKE results, the magnetization $\langle M \rangle$ and the extracted amplitudes $\langle A_1 \rangle$, $\langle A_2 \rangle$ and $\langle A_3 \rangle$ are averaged over the entire probed depth (see text). The dashed yellow line highlights the linear relation of the amplitude $\langle A_3 \rangle$ to the absorbed fluence when the fluence is above the critical fluence.

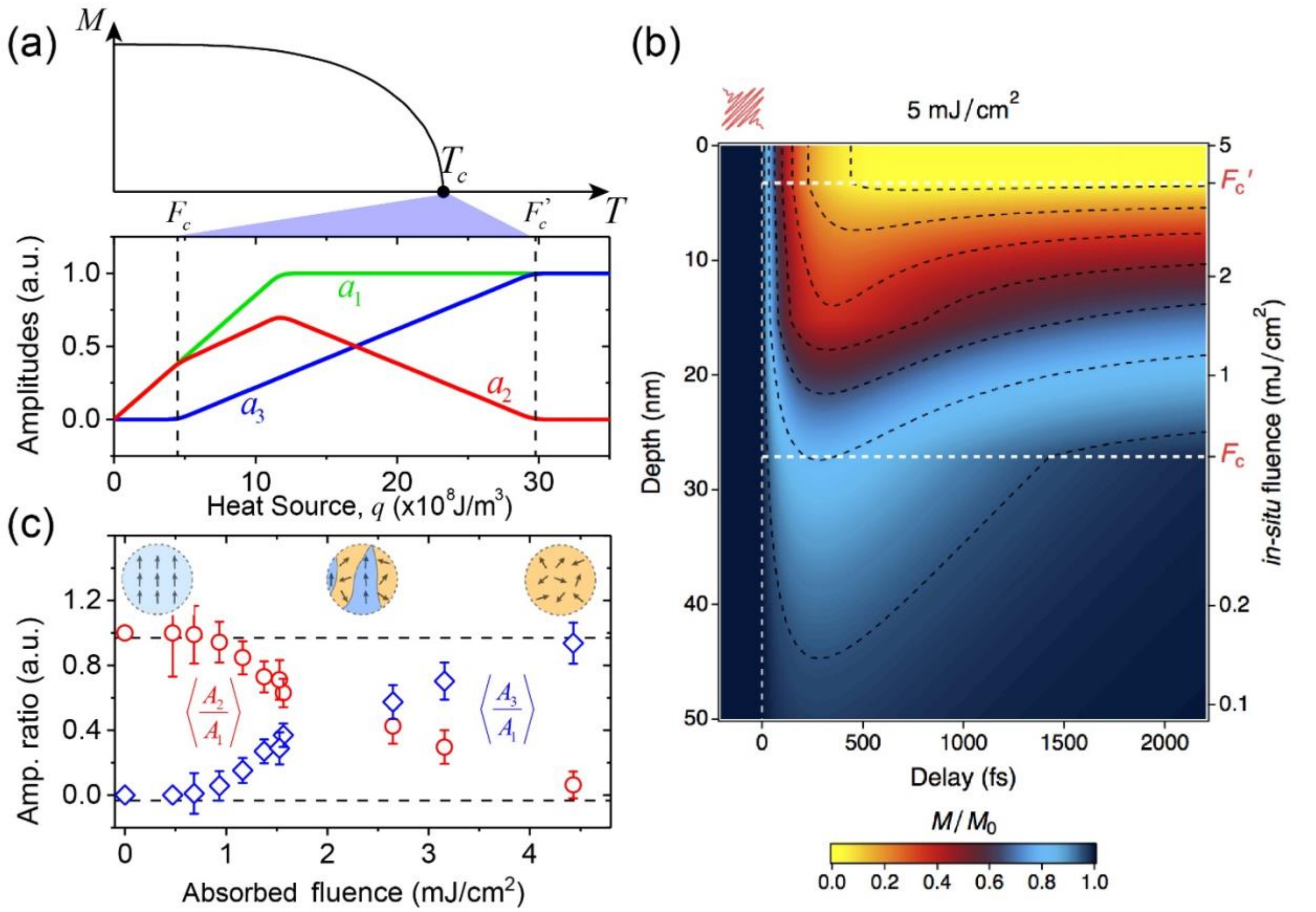

**Figure 4.** (a) Top panel: schematic magnetization of a ferromagnet as a function of temperature under thermal equilibrium with a single critical point ($T_c$). Bottom panel: extracted amplitudes of the change of magnetization in a monolayer of Ni as a function of *in-situ* fluence. The correspondence of $T_c$ to the two critical fluences ($F_c$ and $F_c$') is highlighted. (b) The laser-induced magnetization variation in Ni as a function of time and depth. The black dashed lines represent the contours of equal magnetization. The white dashed lines separate different regions for the in-situ fluence relative to the two critical fluences $F_c$ and $F_c$'. (c) The relative contributions of the fast ($\langle A_2 \rangle$) and slow ($\langle A_3 \rangle$) recovery processes directly extracted from the Tr-TMOKE results in Fig. 3. Inset: potential scenarios for the coexistence of ferromagnetic and paramagnetic phases in different fluence regions.

**Table 1.** Optimum fitting parameters of the Tr-TMOKE results in Fig. 3 to the model, consisting of Eq. (1-4).

| $b_1$ ($cm^2$/mJ) | $b_3$ ($cm^2$/mJ) | $F_c$ (mJ/$cm^2$) |
|---|---|---|
| 0.65±0.01 | 0.31±0.01 | 0.59±0.05 |